\documentclass[amsmath,amssymb,aip,jcp,twocolumn,10pt]{revtex4-1}

\usepackage{dcolumn}
\usepackage{graphicx}
\usepackage{graphics}
\usepackage{setspace}
\usepackage[usenames]{color}

\begin{document}

\title{Correlation correction to configuration interaction singles from
coupled cluster perturbation theory}

\author{Jason N. Byrd}
\author{Victor F. Lotrich}
\author{Rodney J. Bartlett}
\affiliation{Quantum Theory Project, University of Florida, Gainesville, FL 32611}


\begin{abstract}
A new state specific correlation correction to configuration interaction singles
(CIS) excitation energies is preseted using coupled cluster perturbation theory
(CCPT).  
General expressions for CIS-CCPT are derived and expanded explicitly
to first order in the wavefunction and second order in the energy.
By virtue of the nature of CCPT this method is {\it a priori} size
extensive and incorporates infinite order effects into the wavefunction.  This
results in a balanced singles space excited state theory that at
second order is an improvement over the ubiquitous CIS(D) method and
comparable in quality to equation of motion coupled cluster (EOM-CC).
A modest test set composed of the first four excited states from nine small
organic molecules
was used to quantify the accuracy and consistency of the CIS-CCPT2 excitation
energies and density of states.  We find that CIS-CCPT2 has a standard deviation
error of 0.18 eV for excitation energies and 0.14 eV for density of states
compared to EOM-CC, a factor of two better than CIS(D) with a significant
reduction in the maximum deviation as well.
\end{abstract}

\maketitle

\section{Introduction}


The study of electronically excited states of molecules is an important point of
intersection between active experimental investigations and theoretical
methodological progress.  With the ubiquitous use of lasers in optical
spectroscopy the precision obtainable with modern experiments is difficult to
replicate with modern theoretical methods.  Despite this, theory can play an
important role in optical spectroscopy by aiding with excited state assignments,
predicting {\it a priori} the density of states (DOS) that should be expected in
a certain spectrum range and even as a sanity check for the various
complications that can arrise in an experimental apparatus.  

While progress continues in perfecting time dependent density functional theory
(TD-DFT) for use in excited states\cite{dreuw2005} the defacto standards remain
either multireference configuration interaction\cite{szalay2012} (MRCI) or
equation of motion coupled cluster\cite{bartlett2007} (EOM-CC) theory.
Typically truncated to only include single and double excitations,
EOM-CCSD\cite{stanton1993} is qualitatively consistent and approaches
quantitative accuracy in many cases.  Inclusion of triple (either
iteratively\cite{watts1994} or perturbatively\cite{watson2013}) excitations in
the EOM framework generally brings the method into quantitative agreement with
experiment.  However EOM based methods come at a strong computational cost, with
EOM-CCSD having a formal computational scaling of $O(n^6)$ with significant
storage and I/O costs which become demanding for larger molecules.  

Approximations to the excited state problem have been under active research for
many years now, the simplest and oldest being configuration interaction singles (CIS).
Using CIS, the excited state is approximated as a linear combination of single
excitations including the ground state Hartree-Fock wavefunction.  As expected
of a theory that is essentially a mean-field solution to excited states (lacking
all correlation energy as such) the accuracy of CIS is akin to Hartree-Fock
solutions for the ground state.  Early attempts to add correlation to the CIS
wavefunction using second order perturbation theory gave rise to
CIS-MP2\cite{foresman1992} and the more widely used size extensivity corrected
CIS(D)\cite{headgordon1994} method, with various improvements in
accuracy\cite{rhee2007,liu2013,liu2014a} and extensions to higher
orders.\cite{hirata2005} 

In this article, we approach the CIS correlation problem using
coupled cluster perturbation theory\cite{bartlett2010} (CCPT) with the demands
that the resulting method should be {\it a priori} size-extensive, should
contain no cluster opperators of higher particle rank than doubles and include
essential infinite order contributions.  We start in Sec.
\ref{s2} by outlining the general form of single determinant CCPT, followed by
deriving our modification to include the CIS wavefunction into a zero'th order
{\it ansatze} in Sec. \ref{s3}.  The resulting CIS-CCPT energy is explicitly
expanded to second order in Sec. \ref{cisptchap}, with exploratory
numerical calculations presented in Sec. \ref{s5}.

%

\section{\label{s2}Coupled cluster perturbation theory}

Single determinate CCPT\cite{bartlett2010} starts with the exponential
wavefunction {\it ansatz}
\begin{equation}
|\Psi_{\rm CC}\rangle = e^{\hat{T}}|0>
\end{equation}
where the cluster operator,
\begin{equation}
\hat{T} = \hat{T}_1 + \hat{T}_2 + \dots,
\end{equation}
is defined in terms of the excitation operators (here limited to single and
double excitations, with higher order operators defined similarly)
\begin{align}
\hat{T}_1 &= t^{a}_{i} \hat{a}^\dag \hat{i},\\
\hat{T}_2 &= t^{ab}_{ij} \hat{a}^\dag\hat{b}^\dag \hat{i}\hat{j}.\\
\end{align}
Throughout this work the indices $a,b,\dots$ are reserved for particle states
while $i,j,\dots$ refer to hole states.
With this {\it ansatz} the similarity transformed Schr\"{o}dinger equation can be written as
\begin{equation}\label{schrodinger1}
e^{-\hat{T}}\hat{H}e^{\hat{T}}\hat{P} = E \hat{P}
\end{equation}
where $\hat{P}$ is the reference state projection operator,
$\hat{P} = |0\rangle\langle 0|$,
and $\hat{H}$ is the usual normal ordered Hamiltonian
\begin{align}\label{Nhamiltonian}
\hat{H} = 
\hat{P}\hat{H}\hat{P} +
\sum_{pq}f^p_q \lbrace p^\dag q\rbrace_N + 
\frac{1}{4}\sum_{pqrs}\bar{v}^{pq}_{rs} \lbrace p^\dag q^\dag sr\rbrace_N.
\end{align}
Here $f$ is the Fock one-electron matrix, $\bar{v}$ is the antisymmetrized
two-electron integral matrix and $\lbrace\rbrace_N$ denotes normal ordering of
the enclosed operators.
The excitation operator amplitudes $t^{a\dots}_{i\dots}$ are obtained by solving the
system of projected equations
\begin{equation}\label{projected1}
\hat{Q}e^{\hat{-T}}\hat{H} e^{\hat{T}}\hat{P} = 0,
\end{equation}
with $\hat{Q}$ defined to be the operator spanning the complimentary space to
$\hat{P}$,
\begin{equation}
\hat{Q} = |{}^{a}_{i}\rangle\langle{}^{a}_{i}| +
|{}^{ab}_{ij}\rangle\langle{}^{ab}_{ij}| + \cdots.
\end{equation}
For convenience we also can refer to the complimentary space by the particle
excitation rank as $\hat{Q}_1=|{}^{a}_{i}\rangle\langle{}^{a}_{i}|$ for the
single excitation space, $\hat{Q}_2=|{}^{ab}_{ij}\rangle\langle{}^{ab}_{ij}|$
for double excitations and so forth.

Partitioning the Hamiltonian into a zero'th order and perturbation component,
\begin{equation}
\hat{H}=\hat{H}_0+\hat{V}
\end{equation} 
where the exact contribution to $\hat{H}_0$ and $\hat{V}$ is left undefined,
the correlation energy (the zero'th order energy and wavefunction is defined to
be the corresponding mean field solution) can be expressed to order $(m+1)$ as
\begin{equation}\label{en1}
\Delta E^{(m+1)} =
\hat{P}\bar{\bf \cal V}^{(m+1)}\hat{P},
\end{equation}
where
\begin{equation}\label{tranV}
\bar{\bf \cal V}^{(m+1)}
=\left[\hat{V}e^{\hat{T}}\Bigr\rvert^{(m)}\right]_C
\end{equation}
is the similarity transformed perturbation Hamiltonian reduced to only connected
(denoted by the $C$ subscript) terms by application of the
Baker-Campbell-Hausdorff (BCH) identity.
The $(m)$'th order exponential wavefunction is defined here to include all
possible products of $(n)$'th ($n\le m$) order cluster operators (including the $0$'th
order $\hat{1}$ contribution)
\begin{equation}
e^{\hat{T}}\Bigr\rvert^{(m)}=\hat{1}^{(0)} + \sum^\infty_{s=1} \frac{1}{n!}
\left(\hat{T}^{(m/s)}\right)^s \hat{P}.
\end{equation}
The amplitudes of which can be determined by evaluating Eq. \ref{projected1} order by
order using
\begin{equation}\label{wf1}
\hat{Q}\left[(\hat{H}_0+\hat{V})e^{\hat{T}}\Bigr\rvert^{(m)}\right]_C\hat{P} = 0.
\end{equation}

\begin{table}
\caption{\label{hamparttable}Two possible Hamiltonian partitioning schemes.  
Operators denoted with $[n]$ superscripts change particle rank by $n$.}
\begin{ruledtabular}
\begin{tabular}{c|c}
 \multicolumn{2}{c}{Hamiltonian Partitioning} \\
 MBPT & Particle Rank\\
\hline
$\hat{H}_0 = \hat{F}^{[0]} + \hat{F}^{[\pm 1]}$ & $\hat{H}_0 = \hat{F}^{[0]} + \hat{W}^{[0]}$ \\
$\hat{V} = \hat{W}^{[0]} + \hat{W}^{[\pm 1]} + \hat{W}^{[\pm 2]}$ & 
$\hat{V} = \hat{F}^{[\pm 1]} + \hat{W}^{[\pm 1]} + \hat{W}^{[\pm 2]} $\\
\end{tabular}
\end{ruledtabular}
\end{table}

With the CCPT framework so defined, the remaining
choice is how to partition the Hamiltonian.  With future considerations in mind, we
write the normal ordered Hamiltonian (Eq. \ref{Nhamiltonian}) in terms of the one
($\hat{F}$) and two ($\hat{W}$) particle operators, separated by the associated
particle excitation rank:
\begin{equation}
\hat{H} = 
\hat{P}\hat{H}\hat{P} +
\hat{F}^{[0]} + \hat{F}^{[\pm 1]}
+\hat{W}^{[0]} + \hat{W}^{[\pm 1]} + \hat{W}^{[\pm 2]}.
\end{equation}
Using the standard MBPT choice of Hamiltonian partitioning (see Table
\ref{hamparttable}) naturally reproduces the usual MBPT energy and
wavefunction,\cite{shavitt2009} however we are free to choose alternative
partitionings should we desire.  Motivated by the performance of LinearCCD
(coupled cluster doubles truncated to remain linear in $\hat{T}_2$) Bartlett {\it
et al.}\cite{bartlett2010} partition the Hamiltonian in terms of particle
excitation rank.  The effect of this partitioning choice is that any given order in
perturbation theory infinite order contributions are included in the cluster
operators, an important property found in coupled cluster based wavefunctions.

\section{\label{s3}CIS-CCPT(n)}

\begin{table}
\caption{\label{SME}
Excitation energies for a number of small molecules computed with the
aug-cc-pVTZ basis set using MP2/6-31G* geometries.  All units are in electron
volts (eV).}
\begin{ruledtabular}
\footnotesize
\begin{tabular}{ldddd}
& \multicolumn{4}{c}{Excitation Energies}\\
& \multicolumn{1}{c}{CIS}
& \multicolumn{1}{c}{CIS(D)}
& \multicolumn{1}{c}{CIS-CCPT2}
& \multicolumn{1}{c}{EOM-CCSD}
\\
\hline
\multicolumn{5}{l}{Acetone}\\
$A_2$  & 5.16 & 4.40 & 4.20 & 4.44\\
$B_2$  & 8.31 & 5.91 & 6.44 & 6.61\\
$A_1$  & 9.22 & 7.29 & 7.43 & 7.59\\
$A_2$  & 9.20 & 6.84 & 7.52 & 7.64\\
\hline
\multicolumn{5}{l}{Benzene}\\
$B_{3u}$ & 5.99 & 5.28 & 5.20 & 5.16\\
$E_{1g}$ & 6.57 & 6.55 & 6.74 & 6.47\\
$B_{1u}$ & 6.15 & 6.58 & 6.77 & 6.56\\
$A_{2u}$ & 6.99 & 6.59 & 7.31 & 7.10\\
\hline
\multicolumn{5}{l}{Formaldehyde}\\
$A_2$  & 4.51 & 3.96 & 3.72 & 3.94 \\
$B_2$  & 8.63 & 6.62 & 7.16 & 7.24 \\
$B_2$  & 9.43 & 7.53 & 8.07 & 8.11 \\
$A_2$  & 10.08 & 8.03 & 8.66 & 8.66 \\
\hline
\multicolumn{5}{l}{Furan}\\
$A_2$ & 5.96 & 6.12 & 6.28 & 6.12\\
$A_1$\footnote{CIS overlap $\delta S\sim 50$\%.}
      & 7.87 & 6.92 & 6.52 & 6.77\\
$B_2$ & 6.25 & 6.51 & 6.67 & 6.43\\
$B_1$ & 6.46 & 6.64 & 6.85 & 6.67\\
\hline
\multicolumn{5}{l}{Naphthalene}\\
$B_{3u}$ & 5.15 & 4.43 & 4.30 & 4.38\\
$B_{2u}$ & 4.99 & 5.09 & 5.30 & 4.99\\
$A_{u}$  & 5.60 & 5.65 & 5.83 & 5.68\\
$B_{2g}$ & 5.96 & 6.02 & 6.25 & 6.07\\
\hline
\multicolumn{5}{l}{Norbornadiene}\\
$A_2$ & 5.52 & 5.50 & 5.72 & 5.52\\
$B_1$ & 6.06 & 5.88 & 6.16 & 6.06\\
$A_1$ & 6.42 & 6.32 & 6.66 & 6.42\\
$A_2$ & 6.66 & 6.38 & 6.74 & 6.51\\
\hline
\multicolumn{5}{l}{Propanamide}\\
$A''$ & 6.64 & 5.54 & 5.47 & 5.69\\
$A'$  & 8.65 & 6.32 & 6.44 & 6.69\\
$A''$ & 7.57 & 6.42 & 6.66 & 6.65\\
$A'$  & 8.48 & 6.49 & 7.09 & 7.24\\
\hline
\multicolumn{5}{l}{Pyrazine}\\
$B_{1u}$  & 5.10 & 4.36 & 4.24 & 4.31\\
$B_{3u}$  & 5.84 & 5.18 & 5.08 & 5.09\\
$A_u$     & 6.98 & 4.92 & 5.10 & 5.24\\
$B_{3g}$  & 6.69 & 6.15 & 6.12 & 5.93\\
\hline
\multicolumn{1}{l}{MD}   &     & 0.80 & 0.31 & \\
\multicolumn{1}{l}{MAD}  &     & 0.23 & 0.16 & \\
\multicolumn{1}{l}{RMSD} &     & 0.33 & 0.18 & \\
\hline
 & \multicolumn{4}{c}{Relative Density of States Statistics} \\
MD   &      & 0.66 & 0.39 \\
MAD  &      & 0.19 & 0.10\\
RMSD &      & 0.27 & 0.14\\
\end{tabular}
\end{ruledtabular}
\end{table}

To include the CIS wavefunction in the CCPT framework we redefine our zero'th
order wavefunction to be
\begin{equation}\label{ciswf}
|\psi_0^{k}\rangle = \left[e^{\hat{C}^k_1}\right]_{CIS}|0\rangle 
\equiv (1+\hat{C}^k_1)|0\rangle,
\end{equation}
where the $CIS$ subscript denotes that only terms from the exponential up to
linear order (as indicated in Eq. \ref{ciswf}) are retained and
\begin{equation}
\hat{C}^k_1 = c(k)^a_i \hat{a}^\dag \hat{i}
\end{equation}
is a single excitation operator with amplitudes obtained using the
configuration interaction singles (CIS) method\cite{forseman1992} for the $k$'th
state.  Using Eq. \ref{ciswf} the projection operator becomes
\begin{equation}
\hat{P} = (1+\hat{C}^k_1)|0\rangle\langle 0|(1+\hat{C}^{k\dag}_1).
\end{equation}
With this definition of the starting wavefunction the $(0)$'th order reference
energy (given by $\hat{P}\hat{H}_0\hat{P}$) becomes shifted by the $CIS$
excitation energy
\begin{equation}
\omega^k_{CIS} = \hat{P}_0\hat{C}^{k\dag}_1 \hat{H}_0 \hat{C}^{k}_1\hat{P}_0.
\end{equation}
We continue to use the ground state projection operator defined as
$\hat{P}_0=|0\rangle\langle 0|$, but note
that $\hat{Q}$ must be altered to remain orthogonal to $|\psi_0^{k}\rangle$.
Substitution of Eq. \ref{ciswf} into Eq. \ref{en1} gives
\begin{equation}
\label{en2_1}
\Delta E^{k(m+1)}
=
\frac{
\hat{P}_0\left[e^{\hat{C}^{k\dag}_1} 
\bar{\bf \cal V}^{k(m+1)}
e^{\hat{C}^{k}_1}\right]_{CIS}\hat{P}_0
}{
\hat{P}_0\left[e^{\hat{C}^{k\dag}_1}e^{\hat{C}^{k}_1}\right]_{CIS}\hat{P}_0 
}.
\end{equation}
Subtracting the ground state correlation energy, $E^{(m+1)}_{g}$ (computed using
Eq. \ref{en1}), from Eq. \ref{en2_1} and expanding the denominator to cancel
unlinked terms we obtain the excitation correlation energy
\begin{equation}\label{cisptE}
\Delta\omega^{k(m+1)} = 
\hat{P}_0\left[
(1+\hat{C}^{k\dag}_1)
\bar{\bf \cal V}^{k(m+1)}
(1+\hat{C}^{k}_1)
\right]_L\hat{P}_0
-E^{(m+1)}_{g}
\end{equation}
where $L$ denotes that only linked terms are retained.

The associated amplitude projected equations with direct substitution of Eq.
\ref{ciswf} into Eq. \ref{wf1} are
\begin{equation}
\hat{Q}\left[
\left[(\hat{H}_0+\hat{V})e^{\hat{T}^k}\Bigr\rvert^{(m)}\right]_Ce^{\hat{C}^{k}_1}
\right]_{CIS}\hat{P}_0
= 0.
\end{equation}
However this definition does not ensure that all the contributing operators are
connected.  We can strictly enforce connectivity by performing a double
similarity transformation of Eq. \ref{wf1} by premultiplying with
$e^{-\hat{C}^k_1}\rightarrow(1-\hat{C}^k_1)$
before projecting against the complimentary space (an operation we are free to
perform so long as $\hat{Q}$ is orthogonal to
$\hat{P}_0+\hat{C}^{k}_1\hat{C}^{k\dag}_1$), obtaining
\begin{equation}
\label{cisptWF}
\hat{Q}\left[
(\hat{H}_0+\hat{V})e^{\hat{T}^k}\Bigr\rvert^{(m)} (1+\hat{C}^{k}_1)
\right]_C\hat{P}_0
= 0,
\end{equation}
where Eq. \ref{cisptWF} is obtained by repeated use of the BCH identity.

\subsection{\label{cisptchap}Second-order energy}

\begin{table*}
\caption{\label{formaldehyde}Comparison of CIS based perturbative methods for
the problematic case of Formaldehyde.  Results from this work are computed with
the aug-cc-pVTZ basis set while those from Hirata\cite{hirata2005} use the
6-311(2+,2+)G** basis set.  All calculations use the MP2/6-31G* geometry and
units are in electron volts (eV).
}
\begin{ruledtabular}
\begin{tabular}{ldddddddd}
Symmetry & 
\multicolumn{1}{c}{CIS\footnote{This work.}} & 
\multicolumn{1}{c}{CIS(D)\footnotemark[1]} & 
\multicolumn{1}{c}{CIS(3)\footnote{Ref. \onlinecite{hirata2005}.}} & 
\multicolumn{1}{c}{CIS(4)\footnotemark[2]} & 
\multicolumn{1}{c}{MBPT(2)\footnote{Ref. \onlinecite{gwaltney1996}}} & 
\multicolumn{1}{c}{CIS-CCPT2\footnotemark[1]} & 
\multicolumn{1}{c}{EOM-CCSD\footnotemark[1]} & 
\multicolumn{1}{c}{Exp.\footnote{Adopted from Ref. \onlinecite{headgordon1994}}} \\
\hline
$A_2$ & 4.51 & 3.96 & 4.58 & 4.45 & 4.31 & 3.72 & 3.94 & 4.07\\
$B_2$ & 8.63 & 6.62 & 7.82 & 7.39 & 7.10 & 7.16 & 7.24 & 7.11\\
$B_2$ & 9.43 & 7.53 & 8.58 & 8.25 & 7.97 & 8.07 & 8.11 & 7.97\\
$A_2$   & 10.08 & 8.03 & 8.97 & 8.61 & 8.25 & 8.66 & 8.66 & 8.37\\
$A_1$\footnote{CIS overlap $\delta S\sim 50$\%.}
        & 9.67 & 7.82 & 9.24 & 15.71 & 8.02 & 9.02 & 8.22 & 8.14\\
$B_1$   & 9.64 & 9.25 & 9.96 & 9.75 & 9.61 & 9.40 & 9.14 &
\multicolumn{1}{r}{$\cdots$}\\
\end{tabular}
\end{ruledtabular}
\end{table*}

The first order wave function defined by Eq. \ref{cisptWF} is explicitly given as
\begin{equation}\label{1T1}
\hat{Q}_1\left[(\hat{F}^{[0]} + \hat{W}^{[0]}) \hat{T}^{k(1)}_1 + 
\hat{F}^{[+1]}\right]_C\hat{P}_0 = 0 
\end{equation}
\begin{multline}\label{1T2}
\hat{Q}_2\Bigl[
(\hat{F}^{[0]} + \hat{W}^{[0]})(\hat{T}^{k(1)}_1 \hat{C}^k_1 + \hat{T}^{k(1)}_2)
+ \hat{W}^{[+2]} \\
+ (\hat{F}^{[+1]}+\hat{W}^{[+1]})\hat{C}^k_1
\Bigr]_C \hat{P}_0 = 0
\end{multline}
with no higher excitations included at this order due to the connected nature of
Eq. \ref{cisptWF}.  
The second order energy, obtained from Eq. \ref{cisptE} using the amplitudes
defined in Eqs. \ref{1T1} and \ref{1T2}, can be explicitly expressed as
\begin{multline}\label{cisptE1}
\Delta\omega^{k(2)} = 
\hat{P}_0
\Bigl[
(\hat{W}^{[-2]}\hat{T}^{k(1)}_2 + \hat{F}^{[-1]}\hat{T}^{k(1)}_1 - E^{(2)}_g)+  \\
\hat{C}^{k\dag}(\hat{F}^{[-1]} + 
\hat{W}^{[-1]}) \hat{T}^{k(1)}_1 \hat{C}^{k} +
\hat{W}^{[-2]} \hat{T}^{k(1)}_1 \hat{C}^{k}  + \\
\hat{C}^{k\dag}(\hat{F}^{[-1]} + \hat{W}^{[-1]} + \hat{W}^{[-2]} \hat{C}^{k}) \hat{T}^{k(1)}_2
\Bigr]_L
\hat{P}_0.
\end{multline}
Assuming the use of canonical Hartree-Fock starting orbitals ($\hat{T}^{k(1)}_1=0$)
the $T_2$ amplitude equations reduce to
\begin{equation}\label{cisccptT2}
\hat{Q}_2\Bigl[
\hat{W}^{[0]}\hat{T}^{k(1)}_2 + \hat{W}^{[+2]} + \hat{W}^{[+1]}\hat{C}^k_1\Bigr]_C \hat{P}_0 = 0
\end{equation}
and by defining 
\begin{equation}\label{cisccptde}
\delta^{k(2)}_g = 
\hat{P}_0\left[
\hat{W}^{[-2]}\hat{T}^{k(1)}_2
\right]_L\hat{P}_0
- E^{(2)}_g
\end{equation}
along with
\begin{equation}\label{cisccptu}
u^{k(2)} = 
\hat{P}_0\left[
\hat{C}^{k\dag}(\hat{W}^{[-1]} + \hat{W}^{[-2]} \hat{C}^{k}) \hat{T}^{k(1)}_2
\right]_L\hat{P}_0
\end{equation}
the total CIS-CCPT2 excitation energy is compactly expressed as
\begin{equation} \label{cisccpt2E}
\omega^{k(2)} = \omega^k_{CIS} + \delta^{k(2)}_g + u^{k(2)}.
\end{equation}
For explicit spin-orbital equations corresponding to Eqs.
\ref{cisccptT2}-\ref{cisccptu} we refer to Appendix \ref{soeqn}.  

The CIS-CCPT2 equations above introduce infinite order correlation to the CIS
excitation energy root by root with the computational cost of an iterative
$O(n^6)$ scaling.  This computational cost can be contrasted on one hand with EOM-CCSD,
where the scaling limit is also $O(n^6)$ with an additional cost factor of $2$ due to
having both left and right hand vectors.\footnote{Additional storage and computer
I/O scaling costs further penalize EOM-CCSD over CIS-CCPT2.} On the other hand
the prevailing CIS based second order perturbation theory, CIS(D), scales as
$O(n^5)$ but is not {\it a priori} size extensive.  This failing is well known
to plague the CIS(n) perturbation methods,\cite{headgordon1994,hirata2005}
requiring an {\it ad hoc} correction to restore extensivity.  

\section{\label{s4}Electronic Structure Calculations}

\begin{table*}
\caption{\label{acetaldehyde}
Comparison of CIS based perturbative methods for
the problematic case of Acetaldehyde.  Results from this work are computed with
the aug-cc-pVTZ basis set while those from Hirata\cite{hirata2005} use the      
6-311(2+,2+)G** basis set.  All calculations use the MP2/6-31G* geometry and
units are in electron volts (eV).}
\begin{ruledtabular}
\begin{tabular}{ldddddddd}
Symmetry & 
\multicolumn{1}{c}{CIS\footnote{This work.}} & 
\multicolumn{1}{c}{CIS(D)\footnotemark[1]} & 
\multicolumn{1}{c}{CIS(3)\footnote{Ref. \onlinecite{hirata2005}.}} & 
\multicolumn{1}{c}{CIS(4)\footnotemark[2]} & 
\multicolumn{1}{c}{MBPT(2)\footnote{Ref. \onlinecite{gwaltney1996}}} & 
\multicolumn{1}{c}{CIS-CCPT2\footnotemark[1]} & 
\multicolumn{1}{c}{EOM-CCSD\footnotemark[1]} & 
\multicolumn{1}{c}{Exp.\footnote{Adopted from Ref. \onlinecite{headgordon1994}}} \\
\hline
$A''$ & 4.91 & 4.26 & 4.92 & 4.78 & 4.65 & 4.04 & 4.27 & 4.28 \\
$A'$ & 8.51 & 6.34 & 7.47 & 7.10 & 6.88 & 6.83 & 7.00 & 6.82 \\
$A''$ & 9.46 & 7.22 & 8.31 & 7.92 & 7.70 & 7.89 & 7.94 &
\multicolumn{1}{r}{$\cdots$} \\
$A'$\footnote{CIS overlap $\delta S\sim 30$\%.}
     & 9.72 & 8.29 & 8.30 & 4.79 & 7.57 & 8.10 & 7.71 & 7.46 \\
$A'$ & 9.21 & 7.43 & 8.53 & 8.67 & 7.77 & 8.46 & 8.01 & 7.75 \\
$A'$ & 10.33 & 8.13 & 9.16 & 9.25 & 8.42 & 8.48 & 8.75 & 8.43 \\
\end{tabular}
\end{ruledtabular}
\end{table*}

To be consistent with the existing computational excited state
literature\cite{hirata2005,schreiber2008,silva-junior2010} referenced
in this work, molecular equilibrium geometries were obtained using the
MP2/6-31G* level of theory.  All correlation calculations employ
the frozen-core approximation, where the $1s$ orbitals belonging to first row
heavy atoms are excluded from the correlation space.  Electronic structure
calculations were performed on the University of Florida HiPerGator
supercomputer using the massively parallel ACES III\cite{acesiii2008} quantum
chemistry program. 

The use of the TZVP basis set from Sch\"{a}fer {\it et al.}\cite{schafer1992}
was considered here for retaining consistency with existing calculations.
However because of the dependency of our CIS-CCPT2 method on the quality of the
CIS wavefunction, and the tendency of the CIS method to over characterize states
with Rydberg character necessitating the inclusion of diffuse functions in the
basis set, we opted instead to use the significantly more expensive aug-cc-pVTZ
correlation consistent basis set.\cite{kendall1992} Investigative calculations
comparing the TZVP and aug-cc-pVTZ at the EOM-CCSD level of theory show good
agreement for the first excitation energies of the molecules considered in Table
\ref{SME}.  However, the excited state ordering and density of states for the
higher excitation spectrum varied greatly between the two basis sets for a
number of the molecules under consideration.  This is easily attributable to the
better characterization of the excited states through the inclusion of more
angular momentum in the basis set ($5s4p3d2f$ contracted primitive set in the
aug-cc-pVTZ basis while the TZVP basis set is limited to $6s3p1d$ contracted
primitives) in addition to the added diffuse functions.  However this ordering
of states and DOS difference demands great care when comparing with other
published results computed with the smaller TZVP basis set.

\section{\label{s5}Numerical Results and Discussion}

To assess the effectiveness of the CIS-CCPT2 method described here we consider
two numerical metrics: absolute magnitude and ordering of the lowest four
excitation energies and the energy spacing (namely the DOS) relative to
the first excited state.  These metrics were examined using a subset selection
of small molecules from the TBE excited state test
set.\cite{schreiber2008,silva-junior2010}
Additionally we examined in detail two small molecules (Acetaldehyde and
Formaldehyde) known to be problematic for CIS based perturbation
theory.\cite{hirata2005}  All performance statistics (MD for maximum deviation,
MAD for mean absolute deviation and RMSD for root mean square deviation) are
computed relative to the corresponding EOM-CCSD root.  Excited state results
presented here are sorted by the CIS-CCPT2 energy ordering unless stated
otherwise, with the corresponding EOM-CCSD energy assigned by symmetry.


The CIS-CCPT2 predicted excited state spectrum for a selection of small
molecules (a subset of the TBE-1\cite{schreiber2008} test set) is presented in
Table \ref{SME} as well as the corresponding CIS and CIS(D) excitation energy
for that same root.  We find excellent agreement between the
CIS-CCPT2 and EOM-CCSD predicted excitation energies for most of the
molecules examined with an excellent mean deviation of only 0.16 eV.  This can
be compared with CIS(D), the prevalent second-order CIS based perturbation
theory, which has a similarly respectable 0.23 eV average deviation.  Further
comparison shows however that the CIS-CCPT2 method is much more consistent than
CIS(D) with a standard deviation of 0.18 eV compared to 0.33 eV as well as
comparing the maximum deviation, 0.31 eV from CIS-CCPT2 and 0.8 eV for CIS(D).

Additionally we find that the CIS-CCPT2 predicted state ordering agrees exactly
with the EOM-CCSD for most of the molecules listed in Table \ref{SME} with the
exception of Furan and Propanamide.  This is contrasted with the hit or miss
predicted state ordering from CIS(D).  In case of Propanamide the EOM-CCSD state
ordering of $A''\rightarrow A'\rightarrow A''$ is only narrowly determined by an
energy difference of $0.04$ eV between the $aA'$ and $bA''$ states.  Due to this
small energy difference it is difficult to claim that the EOM-CCSD ordering is
indeed definitive.\footnote{Exploratory calculations using EOM-CCSD(T) on these
propanamide states have state orderings that agree with those predicted by
CIS-CCPT2.}   To properly explain the state ordering failure in Furan it is
necessary to examine the overlap of the CIS wavefunction with that of the
associated EOM-CCSD wavefunction, given by
\begin{equation}\label{deltas}
\delta S = 
\frac{\hat{P}_0 L^k_1 C^k_1\hat{P}_0 C^{k\dag}_1 R^{k}_1\hat{P}_0}
{\hat{P}_0 R^{\dag}_1 R^k_1 \hat{P}_0}
\end{equation}
where $L^k_1$ and $R^k_1$ are the left and right hand EOM-CCSD singles
eigenvectors for the $k$'th root.  For simplicity we approximate this overlap by
assuming that $L^k_1\simeq R^{k\dag}_1$.  
The closer
$\delta S$ is to unity the more accurate the CIS wavefunction is as zero'th-order
wavefunction.  For most states investigated here $\delta S>80\%$, however for
the problematic case of the Furan $aA_1$ state this overlap is poor
($\delta S\sim 50\%$), indicative of a stronger doubles contribution to the final
wavefunction. 

We further examine the impact of a poor $\delta S$ by computing the lowest six
roots of Formaldehyde and Acetaldehyde, both molecules known to have states with
significant Rydberg character\cite{gwaltney1995} and poor CIS overlap which is
problematic for CIS based excited state perturbation theory.\cite{hirata2005} It
is evident from Table \ref{formaldehyde} that for Formaldehyde the $aA_1$ state,
with a $\delta S\sim50$\%, is poorly described by either the CIS(n) methods
shown or CIS-CCPT2.
This can be seen again for the miss-ordering of the $bA'$ state of Acetaldehyde,
where the estimated CIS overlap is quite low ($\delta S\sim 30\%$).

\begin{figure*}
\includegraphics[width=2\columnwidth]{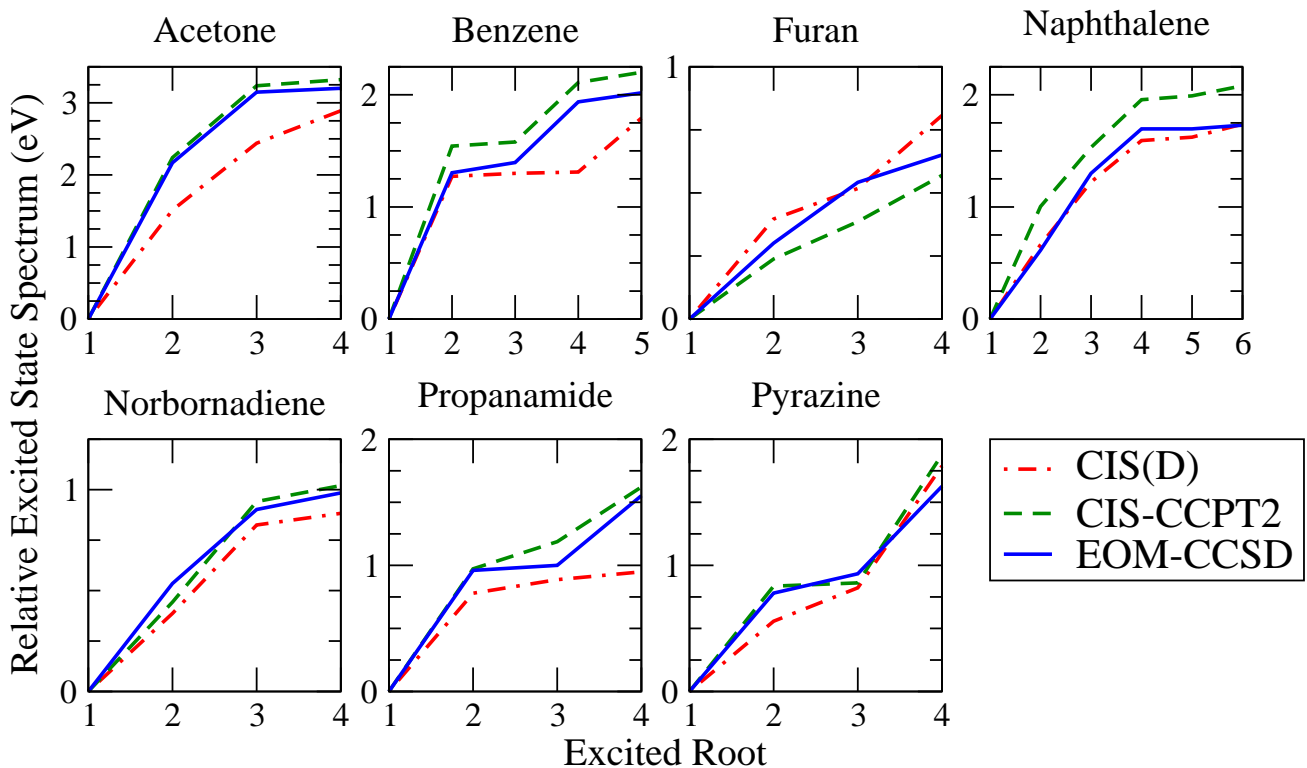}
\caption{\label{SMp} Excited state spectrum relative to the first excited state
for the select set of small molecules investigated here.  Each method's
predicted states are sorted by energy without root by root assignment. Units are
in electron volts (eV).  }
\end{figure*}

In addition to the absolute energy spectrum we also examine the excited DOS
relative to the first excited state, the statistics of which are listed at the
bottom of Table \ref{SME}.  An average deviation of 0.10 eV and a standard
deviation of 0.14 eV for CIS-CCPT2 is quite promising (contrasted with the 0.19
eV average and 0.29 standard deviation for CIS(D)), this is further illustrated
in Fig. \ref{SMp} where the relative energy spectrum for the small molecules
considered here are plotted.  As can be seen, CIS-CCPT2 does well in matching
the features exhibited in the EOM-CCSD excited state spectrum, suggesting that
this method can be accurate as EOM-CCSD for predicting relative energy spacings.
This is of great usefulness in computing solvent related shifts, a topic of
significant interest to us.

\section{Conclusion}

Coupled cluster perturbation theory has been used to derive a correlation
correction to the configuration interaction singles reference wavefunction.  The
resulting CIS-CCPT method is a singles space multi-determinant perturbation
theory with a well defined wavefunction that includes infinite-order effects and
is {\it a priori} size extensive.  The latter being an important consideration
as the usual CIS based perturbation treatments are only size extensive by {\it
ad hoc} omission of certain terms.\cite{headgordon1994,hirata2005}  The first
order wavefunction and second order energy for CIS-CCPT has been explicitly
presented in Section \ref{cisptchap} (see Eqs.  \ref{cisccptT2}-\ref{cisccptu})
and implemented in the ACES III\cite{acesiii2008} quantum chemistry package.

The quality of this CIS-CCPT2 method is demonstrated by examining the excited
state spectrum of a number of small molecules (see Table \ref{SME}).  The
resulting average (standard) deviation of 0.16 (0.18) eV as
compared to EOM-CCSD is a marked improvement over CIS(D) with an average
(standard) deviation of 0.23 (0.33) eV.  Additionally, the density of states
relative to the first excited state also compares well with EOM-CCSD, with an
average (standard) deviation of 0.10 (0.14) eV.  To illustrate this last
comparison we plot in Fig. \ref{SMp} the relative excited state spectrum of the molecules
considered here, where CIS-CCPT2 can be seen to trace the features seen in the
EOM-CCSD excited spectrum.  It can also be noted that the CIS-CCPT2 and CIS(D)
DOS generally bracket the corresponding EOM-CCSD spectrum, a convenient trend for
identifying problematic states.  To further investigate the behavior of
CIS-CCPT2 for difficult cases we compare the first six excited states of
Formaldehyde and Acetaldehyde (known to be problematic for CIS based
perturbation theory\cite{hirata2005}) to both EOM-CCSD and experimental results.
We find that CIS-CCPT2 has excellent agreement with the comparison values except
in the expected situation where the CIS wavefunction overlap with the EOM-CCSD singles
vector (Eq. \ref{deltas}) is poor.

\section{Acknowledgments}

The authors would like to acknowledge funding support from the United States
Air Force Office of Scientific Research.

\appendix
\section{\label{soeqn}Spin-Orbital Equations for CIS-CCPT2}
Explicit spin-orbital equations that define the CIS-CCPT2 amplitude and
energy expressions are presented below.   
The amplitude Eq. \ref{cisccptT2} is (assuming Einstein notation)
\begin{multline}\label{1T2so}
\bar{v}^{ab}_{ij}
-\epsilon^{ab}_{ij}t(k)^{ab(1)}_{ij}
+ \frac{1}{2} \bar{v}^{ab}_{cd} t(k)^{cd(1)}_{ij}
+\frac{1}{2} \bar{v}^{ij}_{ml} t(k)^{ab(1)}_{ml} \\
+ P(ij)P(ab)\bar{v}^{mb}_{cj}t(k)^{ac(1)}_{im}
+ P(ab)\bar{v}^{mb}_{ij} c(k)^a_m
=0
\end{multline}
where
\begin{equation}
\bar{v}^{pq}_{rs} = \langle p_1r_1||q_2s_2\rangle
\end{equation}
are antisymmetric two electron integrals,
\begin{equation}
\epsilon^{ab}_{ij} = ( f^i_i + f^j_j - f^a_a - f^b_b)
\end{equation} are
the Hartree-Fock orbital energies and 
\begin{equation}
P(ij)u_{ij} = u_{ij} - u_{ji}
\end{equation}
is the antisymmetrized permutation operator.  
From the CIS-CCPT2 energy expression (Eq. \ref{cisccpt2E}) the
the $\delta^{k(2)}_g$ contribution (Eq. \ref{cisccptde}) is
\begin{equation}
\delta^{k(2)}_g = 
\frac{1}{4}\bar{v}^{ij}_{ab}t(k)^{ab(1)}_{ij} - E^{(2)}_g
\end{equation}
while  $u^{k(2)}$ (Eq. \ref{cisccptu}) is given by 
\begin{multline}
u^{k(2)} = 
\frac{1}{4}c(k)^a_m \bar{v}^{mb}_{ij} t(k)^{ab}_{ij}\\
+\frac{1}{2} c(k)^a_i \bar{v}^{bc}_{jm} (
t(k)^{ca}_{jm} c(k)^b_i +  
t(k)^{cb}_{im} c(k)^a_j + 
2 t(k)^{ac}_{im} c(k)^b_j)
\end{multline}

\section{\label{appb}Connection to CIS(n) perturbation theory}

A satisfying connection to CIS(n) perturbation theory can be drawn from the
CIS-CCPT2 equations with some judicious truncation and approximation.  We start by approximating
the $\hat{T}_2$ equation from Eq. \ref{1T2} by the zero'th iteration
approximation
\begin{equation}
\hat{Q}_2\hat{\cal T}^{k(1)}_2\hat{P}_0 = 
\hat{R}_0\left[
\hat{W}^{[+2]}+
\hat{W}^{[+1]}\hat{C}^k_1
\right]\hat{P}_0
\end{equation}
using the denominator shifted resolvent (infinite order methods like
CIS-CCPT(n) are of course invariant to denominator shifts)
\begin{equation}
\hat{R}_0 = 
\hat{Q}
\left[
\frac{1}{E^{k(0)}-F^{[0]}}
\right]
\hat{Q}
= \frac{\hat{Q}}{\omega^k_{CIS} + \epsilon^{ab\dots}_{ij\dots}}.  
\end{equation}
With this approximation Eq. \ref{cisccptde} is reduced to
\begin{equation}
\delta^{k(2)}_g \simeq
\hat{P}_0\left[
\hat{W}^{[-2]}(\hat{R}_0\hat{W}^{[+1]}\hat{C}^k_1)
\right]_L\hat{P}_0
\end{equation}
which is generally small and so is neglected,
while Eq. \ref{cisccptu} becomes
\begin{multline}\label{appb1}
u^{k(2)} \simeq
\hat{P}_0\Bigl[
(\hat{C}^{k\dag}\hat{W}^{[-1]} + \hat{W}^{[-2]} \hat{C}^{k})\times \\
((\hat{R}_0\hat{W}^{[+2]})+(\hat{R}_0\hat{W}^{[+1]}\hat{C}^k_1))
\Bigr]_L\hat{P}_0.
\end{multline}
To finish the {\it ad hoc} link to the second order CIS(D) equations it is
necessary to realize that in the CIS(n) derivation does not start with the
$(1+\hat{C}^k_1)$ ansatz but instead the fundamentally different
$(\hat{C}^k_1)$ starting wavefunction.  This means that in the CIS(n) expansion
only two of the four terms\footnote{These two surviving terms are generally of
the same magnitude and opposite sign.} in Eq. \ref{appb1} can occur.  Simplification of Eq.
\ref{appb1} directly gives the CIS(D) energy equation:
\begin{multline}\label{cispd}
u^{k(2)} \simeq 
\hat{P}_0\Bigl[
\hat{C}^{k\dag} \hat{W}^{[-1]} (\hat{R}_0\hat{W}^{[+1]}\hat{C}^k_1) +  \\
\hat{C}^{k\dag} \hat{W}^{[-2]} \hat{C}^k_1 (\hat{R}_0\hat{W}^{[+2]}) 
\Bigr]_L\hat{P}_0.
\end{multline}


%

\end{document}